\documentclass{amsart}       
\usepackage{graphicx}
\usepackage{mathptmx}      
\usepackage{latexsym}
\usepackage{amsmath}
\usepackage{epsfig}
\usepackage{amssymb}
\usepackage{enumerate}
\usepackage{xcolor}

\usepackage{amsmath, amsfonts, amssymb} 
\usepackage{mathtools}
\usepackage{enumitem}
\usepackage[numbers, sort&compress]{natbib}
\usepackage[colorlinks]{hyperref}
\hypersetup{
	citecolor=blue,
	linkcolor=red,
}

\title{SPRINT: A fast, new software tool for reconstructing the evolutionary past of polyploid datasets}

\author{Liam J. Maher, Taoyang Wu and Katharina T. Huber}
\address{School of Computing Sciences, University of East Anglia,
	Norwich, UK }
\email{L.Maher@uea.ac.uk}
\address{School of Computing Sciences, University of East Anglia,
	Norwich, UK }
\email{Taoyang.wu@uea.ac.uk}
\address{School of Computing Sciences, University of East Anglia,
	Norwich, UK}
\email{K.Huber@uea.ac.uk}



\date{\today}

\begin{document}

\begin{abstract}
Polyploidization is an important evolutionary process which affects organisms ranging from plants to fish and fungi. The signal left behind by it is  in the form of  a species'  \textit{ploidy level} (number of complete chromosome sets found in a cell) which is inherently non-treelike.  Currently available tools for reconstructing the evolutionary past of a polyploid dataset generally start with a multi-labelled tree obtained for a dataset of interest and then derive a (phylogenetic) network from that tree in some way that reflects that past by interpreting the networks's  vertices of indegree at least two as polyploidization events. Since obtaining such a tree can be computationally expensive it is paramount to have alternative approaches available that allow one to shed light into the reticulate evolutionary past of a  polyploid dataset. SPRINT aims to reconstruct the evolutionary past of a polyploid dataset in terms of a  binary network which realises the dataset's \textit{ploidy profile} (vector of ploidy levels of the dataset's taxa) and requires the fewest number of polyploidization events. It does this by 
representing the  ploidy level of a species $x$ in terms of the number of directed paths from the root of the network to the leaf of the network labelled by $x$. SPRINT is distributed on GitHub: https://github.com/lmaher1/SPRINT.
		
\end{abstract}
	
	\maketitle

\section{Introduction}
Polyploidization is an important evolutionary driver which affects organisms ranging from plants to fish and fungi. Existing tools for reconstructing the evolutionary past of polyploid datasets include those described  in \cite{HOLM06,LSHPOM09}. Their main drawback is their reliance  on the  availability of a multi-labelled tree for  the species of interest (see e.g.\,\cite{HOLM06}) because  such a tree might not be readily available. 
Tools that circumvent this problem are currently largely lacking thereby hindering the exploitation of polyploidy information by evolutionary biologists. Here, we present the {\it \textbf{S}pecies \textbf{P}loidy \textbf{R}ealisation of \textbf{I}ntegers with \textbf{N}etworks \textbf{T}ool} (SPRINT) software tool which solely relies on a dataset's ploidy levels for shedding light into its evolutionary past. 
\section{SPRINT} 
Given a polyploid dataset SPRINT uses a heuristic approach to reconstruct the evolutionary past of a polypoid dataset in terms of a network that realizes its ploidy profile.  In view of \cite{HM22}, the number of polyploidization events required by the inferred network is minimal under certain conditions (see below) and an upper bound in general. The generated network may be viewed within SPRINT's Graphical User Interface (GUI) or directly exported in DOT format to a graph drawing tool of choice (e.g.\,\cite{SMOBWRASI03}).

The algorithm underlying SPRINT is best described as a two-step process (see \cite{HM22,HM222} for details and Section~\ref{example} for an example). Starting with a ploidy profile $\vec{m}_1=\vec{m}$, the algorithm first constructs a {\em simplification sequence} $\sigma(\vec{m})$ for 
$\vec{m}$. This is a sequence of ploidy profiles $m_i$, $i\geq 1$, where $\vec{m}_{i+1}$ is obtained from $\vec{m}_i$ by repeatedly taking the difference $\alpha=\mu_1-\mu_2$ between the largest ploidy level $\mu_1$ in $\vec{m_i}$  and the second largest ploidy level $\mu_2$ in $\vec{m}_i$ until a ploidy profile of the form $(k)$ or $(k,1,\ldots, 1)$, $k$ a positive integer, is obtained.  This difference retains key properties of $\vec{m}$ and  the following three cases are distinguished : A) $\alpha=0$, B) $\alpha>\mu_2$, or C)  $\alpha\leq \mu_2$. 
The second step is  a {\it traceback} of $\sigma(\vec{m})$ which applies at each step a certain network operation
that  in some sense undoes the corresponding case in the construction of $\sigma(\vec{m})$. By construction, the resulting network realizes the corresponding ploidy profile in $\sigma(\vec{m})$ at each traceback step. These operations are related to the so called cherry reduction operations for networks introduced in \cite{IJJM21} (see also \cite{HM222}). 

SPRINT is guaranteed to find an optimal phylogenetic network that realizes a plody profile $\vec{m}$ if  the following two rules are not violated: (i) the traceback  is initialized  with an optimal network for the terminal element of $\sigma(\vec{m})$ (see \cite{sprint-webpage} for an example of a non-optimal initialisation for the ploidy profile (47)); (ii) the traceback step that corresponds to operation B must not be used twice in a row. If it has been applied twice in a row, SPRINT terminates and outputs the message "no optimal network found".

Two methods are implemented in SPRINT to initialize
the construction of the aimed for optimal network, namely the {\it Binary representation} and {\it Prime factor decomposition} methods \cite{HM22}. By default, the network that requires the minimum number of polyploidization events is chosen (prime factor decomposition is chosen in the case of a tie). However, a choice between these two methods is permitted by SPRINT, as well as the ability to add a custom initialisation network should it be favoured e.g. in the case of the ploidy profile (47) -- see above. 
\section{Example}
\label{example}
\begin{figure}[h]
\centering
\includegraphics[scale=0.34]{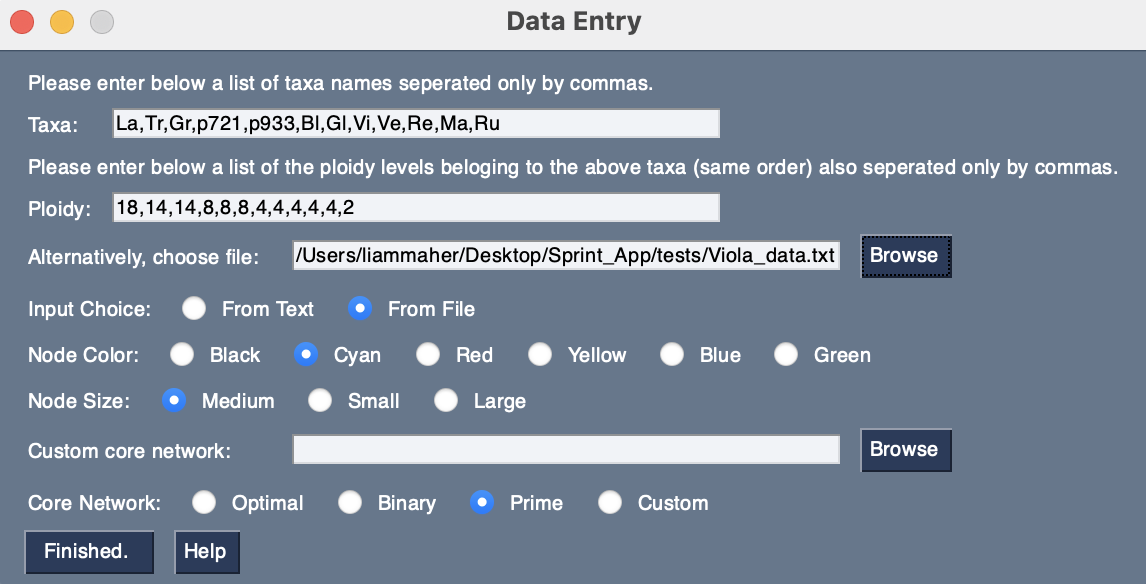}
\caption{SPRINT's input GUI (including guidance) for the ploidy profile 
$\vec{m}$ in Section~\ref{example}.}
\label{fig:input}
\end{figure}

\begin{figure}[]
\centering
\includegraphics[scale=0.39]{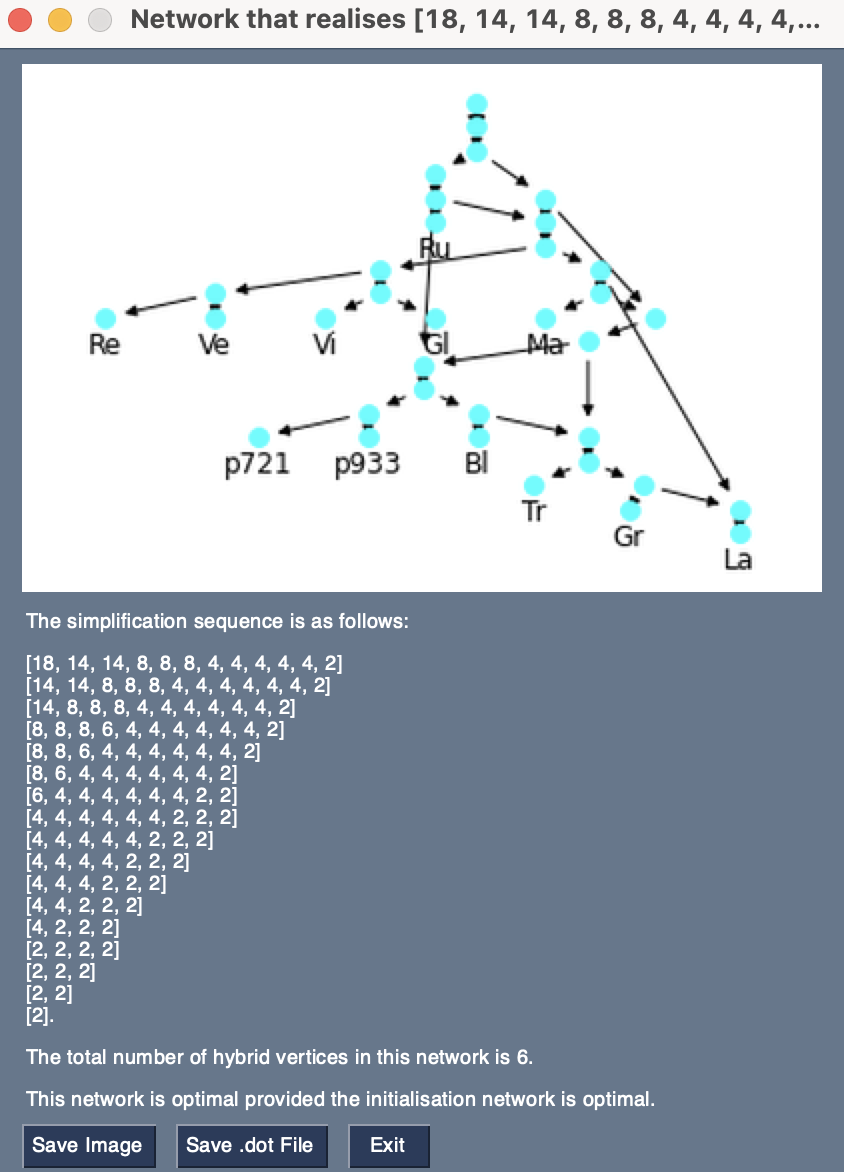}
\caption{SPRINT's output GUI for the ploidy profile $\vec{m}$ from Section 3. From top to bottom, the generated netwok, the ploidy profiles in $\sigma(\vec{m})$ as well as the total number of polyploidization events required by the network.}
\label{fig:output}
\end{figure}

Consider the ploidy profile $\vec{m}=(18,14,14,8,8,8,$
$4,4,4,4,4,2)$ where the components give the ploidy level of the species in the (ordered) Viola data set $\{La,Gr,Tr,$ $Bl,p933,p721,Ma,Gl,Vi,Ve,Re, 
Ru\}$ considered in \cite{HM22}. So, for example, species $La$ indexes the first component of $\vec{m}$ and so has ploidy level 18. Then a screenshot for how to input this dataset into SPRINT's GUI is presented in Fig.~\ref{fig:input}.
A screenshot of SPTINT's output GUI that shows the network generated for $\vec{m}$ as well as $\sigma(\vec{m})$ and the required number of polyploidization events is presented in Fig.~\ref{fig:output}. It should be noted that the network is optimal in this case (see \cite{HM22}).
\section{Implementation and Testing}
SPRINT is an open source Python software tool which works on any platform where Python 3.9 or above and the NetworkX package is installed. SPRINT comes with a GUI for ease of data entry and presentation of output. 
Data entry via a .txt file using SPRINT's file selection option is also offered.  The color and size of vertices in the generated network are adjustable and all graphics can be saved directly in .png format for use in publications. Although the authors suggest using pygraphviz for the DOT file and visualisation of output (default),  an alternative is provided 
in the form of SPRINT-nxspring.py.
 In addition to SPRINT, a manual, and test files may be found on SPRINT's GitHub site. 
SPRINT was rigorously tested with ploidy profiles containing over 50 randomly chosen ploidy levels ranging from 1 to 30 which it completed in under 5 seconds each time.
\section{Discussion}
We have developed the network reconstruction tool SPRINT which is written in Python 3.9. SPRINT uses a heuristic approach and aims to compute a network that requires the minimum number of polyploidization events  to explain a dataset's polyploid profile. SPRINT's output can either be directly cut and paste into a publication or can be input into other network drawing programs  (e.g.\,\cite{SMOBWRASI03}).  Future work might include  combining SPRINT with existing programs such as those in \cite{HOLM06,LSHPOM09} to help, for example,  inform the generation of multi-labelled trees or measure how different networks are that realize the same ploidy profile.

\section*{Acknowledgements and Funding}
{\em Conflict of interest:} none declared. 
{\em Funding:} not applicable.

 \bibliographystyle{plain}
 \bibliography{bibliography}

\end{document}